# Strain-induced programmable half-metal and spin-gapless semiconductor in an edge-doped boron nitride nanoribbon

Shuze Zhu[1], Teng Li[1]

Department of Mechanical Engineering, University of Maryland, College Park, MD 20742

**Abstract:** The search for half-metals and spin-gapless semiconductors has attracted extensive attention in material design for spintronics. Existing progress in such a search often requires peculiar atomistic lattice configuration and also lacks active control of the resulting electronic properties. Here we reveal that a boron-nitride nanoribbon with a carbon-doped edge can be made a half-metal or a spin-gapless semiconductor in a programmable fashion. The mechanical strain serves as the on/off switches for functions of half-metal and spin-gapless semiconductor to occur. Our findings shed light on how the edge doping combined with strain engineering can affect electronic properties of two-dimensional materials.

**Keywords:** Edge Doping; Boron-Nitride; Strain Engineering; Half-Metal; Spin-Gapless Semiconductor;

---

[1] Electronic mail: LiT@umd.edu (Teng Li); shuzezhu@umd.edu (Shuze Zhu)

## I. Introduction

Spintronics have been proposed as a new approach that can revolutionize electronic devices [1]. Spintronic devices require spin currents, which use not only the electron charge, but also the spin as transport property [1]. Therefore the realization of spintronics hinges upon finding new types of materials that are spin-polarized. Half-metals (HM) have been long sought as candidates for spintronics applications due to their exceptional electronic structure [1]. Recently, inspired by zero-gap semiconductors, a new concept of spin-gapless semiconductors (SGS) is proposed [2]. The most striking advantage of SGS is that the excited charge carriers can be 100% spin-polarized near Fermi level, highly desirable for spintronic materials [3]. In general, HM and SGS function permanently with respect to a specific atomistic lattice configuration (e.g., Heusler compounds [4,5]), the searching of which has attracted substantial efforts [6-10]. Despite of existing experimental demonstrations of HM and SGS, it remains as a grand challenge to achieve active control of these functions, a highly desirable feature to enable unconventional design of spintronics devices. In this letter, using density functional theory (DFT) calculations, we reveal programmable HM and SGS in a zigzag boron nitride nanoribbon with a carbon-doped edge. We show that a mechanical strain can serve as the switch to turn on and off the HM and SGS functions. Our findings shed light on fertile opportunities toward active control of HM and SGS functions via strain engineering of two-dimensional (2D) materials.

Hexagonal boron nitride (h-BN) is a non-centrosymmetric dielectric 2D crystal. The structural asymmetry of h-BN introduces spontaneous polarization with a residual dipole moment [11]. Recently, hetero-epitaxial monolayers composed of graphene and h-BN (C/BN) have been demonstrated experimentally [12,13], opening up new possibilities on tailoring the electronic structure of such hetero-epitaxial materials. For example, several theoretical works have shown that half-metallic properties can be achieved in hybrid C/BN monolayers [8,9,14,15]. The key feature of a half-metal is that in one spin channel, the Fermi level passes through its conduction band, while in another spin channel, it is still semi-conductive [Fig. 1(a)]. In those studies, the transition from a semiconductor to a half-metal is usually permanently coupled with the geometry of the hybrid. For example, the width of zigzag graphene nanoribbon dictates whether the hybrid structure is metallic or not [8]. Nevertheless, programmable control of such a transition is yet to be achieved so far.

Graphene is a promising spintronics material due to its room-temperature spin transport with long spin-diffusion lengths [16,17]. Graphene has a particular band structure in which the edges of valence and conduction bands just touch at the Fermi level, representing a class of solids called the zero-gap semiconductors [18,19]. For zero-gap semiconductors, no threshold energy is required to move electrons from occupied states to empty states. Therefore, such a unique band structure renders graphene with exceptional electronic properties, such as high electron mobility [20], quantum hall effect [21] and size-dependent semiconductivity [22]. Inspired by zero-gap semiconductors, SGS is proposed, with the defining feature that at least in one spin channel, the Fermi level falls within a zero-width gap [Fig. 1(b)]. The intensive search for SGS has led to several candidate material systems, such as doped $PbPdO_2$ [2], N-doped zigzag graphene nanoribbons [6], h-BN nanoribbons with vacancies [23]. However, the permanent SGS function of such materials is dictated by their peculiar atomistic lattice configurations and thus lacks tunability.

The properties of 2D crystals are strongly tied to their lattice structure [24]. Both graphene and h-BN have remarkable flexibility and can sustain large elastic strains. These features suggest rich opportunities of tuning electronic properties of 2D crystals via strain engineering [11,25]. So far, explorations of such opportunities have been mainly focused on monolithic 2D crystals. Strain engineering of hetero-epitaxial 2D materials (e.g., hybrid C/BN [26]) holds the potential to reveal unconventional electronic properties that are otherwise non-existing in their individual constituents, a topic remaining largely unexplored so far. In this letter, we show that a mechanical strain applied on a C/BN monolayer nanoribbon can trigger a semiconductor-to-HM transition or a semiconductor-to-SGS transition, depending on the lattice structure of the nanoribbon. Such transitions are rather sharp with respect to the applied mechanical strain, promising their possible applications as switches in spintronics, a highly desirable but hard-to-achieve device function.

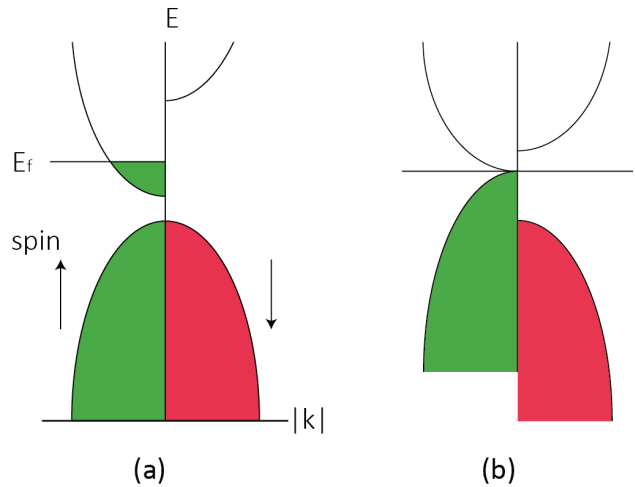

FIG. 1. Electronic band structures of a half-metal (a) and a spin-gapless semiconductor (b). Figures show the energy-momentum (E-k) dispersions with respect to the Fermi level ($E_f$). The green and red areas indicate the spin-up and spin-down states, respectively.

## II. Methodology

We consider C/BN nanoribbons with zigzag edges in two distinct lattice structures [Fig. 2(a) and Fig. 2(c)]. In the first lattice structure [Fig. 2(a), referred to as Lattice #1 hereafter], one zigzag edge is made of nitrogen atoms (in blue) and another edge is doped with a single row of hexagonal carbon rings (in grey). Both nitrogen and carbon atoms along the two edges are hydrogen-saturated. The second lattice structure [Fig. 2(c), referred to as Lattice #2 hereafter] can be constructed by switching the nitrogen (in blue) and boron (in gold) atoms in Lattice #1, so that its one edge is boron-terminated and another edge is still carbon-doped in the same was as in Lattice #1. C/BN nanoribbons with various widths are considered. The nanoribbon width is characterized by the number of full hexagonal boron-nitride rings (BNRs) in its width direction [labeled as BNR1, BNR2, etc., Fig. 2(a) and Fig. 2(c)]. The C/BN nanorribon is subject to a uniaxial tensile strain along its length direction.

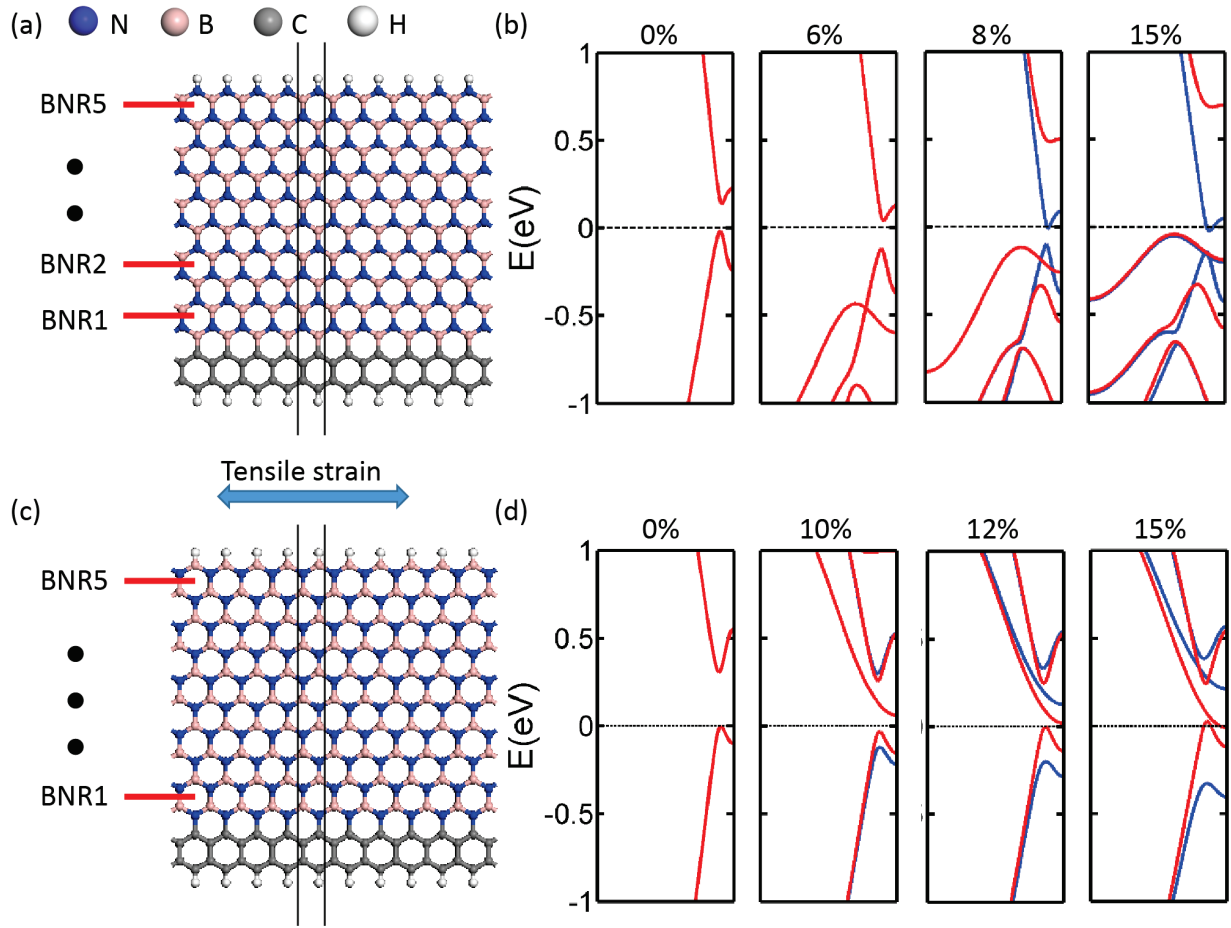

FIG. 2. (a) and (c) show two distinct lattice structures of hybrid C/BN nanoribbons, referred to as Lattice #1 and Lattice #2, respectively. A tensile strain is applied along the nanoribbon length direction. The atoms demarcated between the two solid lines define the supercell configuration for DFT calculation. (b) and (d) plot band structures of (a) and (c), respectively, in undeformed state (0%) and under various tensile strains. The Fermi energy is shifted to zero energy level. The blue and red line are for spin-up and spin-down states, respectively. Here we consider C/BN nanoribbons with 5 full BNRs in their width direction (i.e., BNR5)

We perform first-principle DFT calculations in a supercell configuration [demarcated between two solid lines in Fig. 2(a) and Fig. 2(c)] by utilizing the SIESTA code [27]. The spin-polarized generalized gradient approximation in the framework of Perdew-Burke-Ernzerhof is adopted for the exchange-correlation potential. Numerical atomic orbitals with double zeta plus polarization are used for basis set, with a plane-wave energy cutoff of 300 Ry. At least 50 (along the nanoribbon axis direction) $\times$ 3 $\times$ 1 Monkhorst-Pack k-points are set. We introduce vacuum separation of 100 Å for out-of-plane direction and at least 50 Å for in-plane direction so that the nanoribbon is rather a one-dimensional isolated system. Self-consistent field tolerance is set to $10^{-6}$. Geometry optimizations are first performed before the calculation of band structure and density of states.

**III. Results and discussion**

Figure 2(b) shows the band structure of a hybrid C/BN nanoribbon in Lattice #1 under various tensile strains. Here the width of the nanoribbon is BNR5. The undeformed C/BN nanoribbon has a direct band gap of about 0.15 eV. Such a direct band gap remains nearly unchanged when the nanoribbon is subject to a modest tensile strain (up to ~6%). The spin-up and spin-down band lines overlap with each other. However, when the applied tensile strain increases to 8%, the spin-up (blue) and spin-down (red) band lines separate. The spin-up state (blue) are metallic as the Fermi level passes through its conduction band, while the spin-down state (red) still remains semi-conductive, rendering the nanoribbon under tension as half-metallic. Similar band structures can be observed as the applied tensile strain further increases (e.g., to 15%). In other words, there exists a threshold tensile strain (about 8%), below which a C/BN nanoribbon in Lattice #1 is a semiconductor while above which it becomes a HM. Such a threshold tensile strain is below the elastic limit of the C/BN nanoribbon, so that such a transition is reversible by tuning the tensile strain level.

Interestingly, Fig. 2(d) further reveals a sharp transition of a C/BN nanoribbon in Lattice #2 between semiconductor and SGS as the applied tensile strain varies. The undeformed C/BN nanoribbon in Lattice #2 has a direct band gap of about 0.35 eV. The spin-up and spin-down band lines overlap with each other. However, when the applied tensile strain increases to 10%, the spin-up (blue) and spin-down (red) band lines start to separate. At a tensile strain of 12%, both the highest valence band and the lowest conduction band of the spin-down electrons touch the Fermi level at different points in K space, while the spin-up gap still retains. Such a band structure clearly defines a SGS behavior [2]. Similar SGS band structures are also shown as the applied tensile strain further increases (e.g., 15%). The threshold tensile strain (about 10%) is also below the elastic limit of the C/BN nanoribbons, suggesting the reversibility of the semiconductor-to-SGS transition.

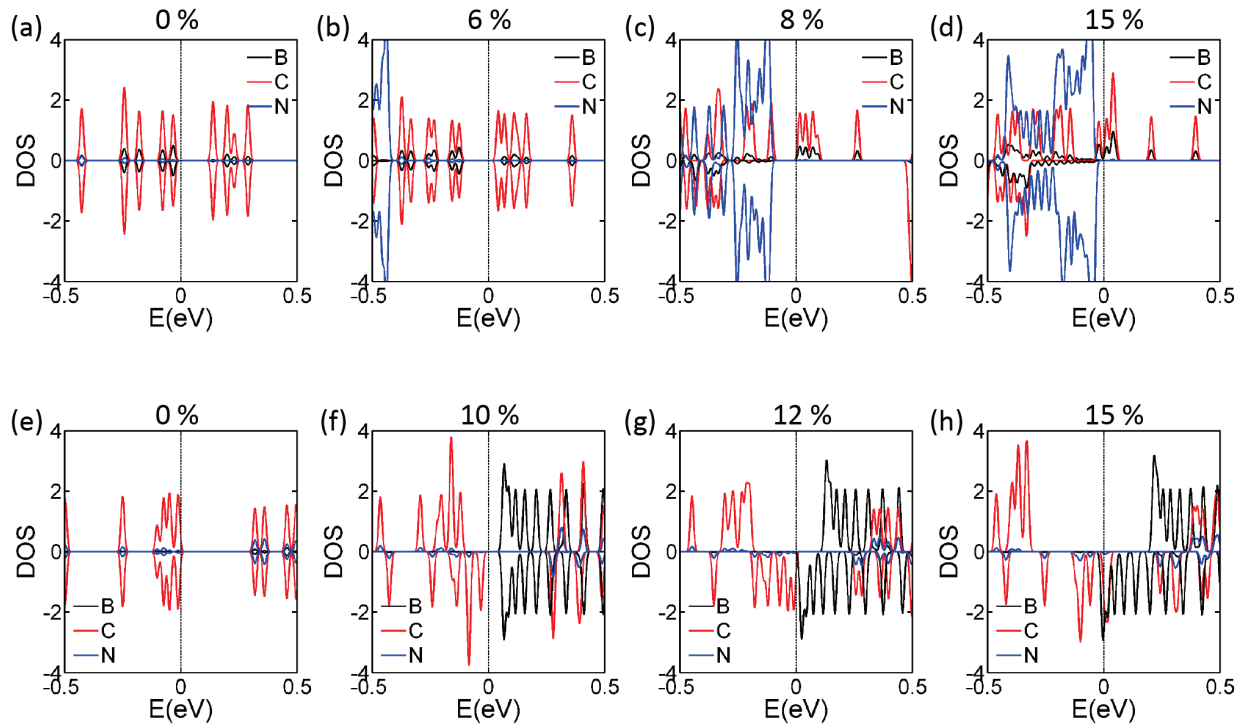

FiG. 3. The electronic density of states (DOS) for the two lattice structures under various tensile strains. Here the projected density of states (PDOS) for 2p orbitals are shown. (a) to (d) correspond to the band structure of Lattice #1 as in Fig. 2(b). (e) to (h) correspond to the band structure of Lattice #2 as in Fig. 2(d).

To shed light on the mechanistic understanding of the formation of HM and SGS in hybrid C/BN nanoribbons, we further calculate the density of states (DOS) for the two lattice structures under various tensile strains. Figures 3(a)-3(d) plot the partial density of states (PDOS) decomposed onto 2p orbitals of each element type for a C/BN nanoribbon in Lattice #1 under various tensile strains corresponding to Fig. 2(b). Positive DOS values concern the spin-up states. In undeformed state [Fig. 3(a)], there are sizeable gaps in both spin-up and spin-down band structures, with the same value about 0.15 eV. For both spin channels, the bottom of the conduction bands mainly originates from the 2p orbitals of carbon atoms (red curves), and so is for the top of the valence bands. As the applied tensile strain increases to 6% [Fig. 3(b)], the bottom of the conduction bands originating from carbon atoms shifts toward the Fermi level, while the top of the valence bands originating from carbon atoms moves away from the Fermi level. As a result, a gap of about 0.15 eV retains. As the tensile strain increases to 8% [Fig. 3(c)], for the spin-up states, the bottoms of the conduction bands from both carbon (red curves) and boron (dark curves) atoms touch the Fermi level. On the other hand, for the spin-down states, the top of the valence bands, which at this strain mainly originates from nitrogen atoms (blue curves), has a sizeable energy gap (about 0.6 eV) to the bottom of the conduction bands originating from carbon atoms. As the applied tensile strain further increases to 15% [Fig. 3(d)], for the spin-up states, the Fermi level passes through the conduction bands from carbon and boron atoms, while the top of the valence bands from nitrogen atoms moves toward the Fermi level in comparison with its position in Fig. 3(c). For the spin-down states, the top of the valence bands from nitrogen atoms also moves towards the Fermi level, but has a large gap to the bottom of the conduction bands. The DOS results in Fig. 3(c) and Fig. 3(d) exhibit the typical HM characteristics. The discussion above also suggests that the contribution from the 2p orbitals of carbon atoms at the ribbon edge is essential to render such HM characteristics in hybrid C/BN nanoribbons under tension.

Figures 3(e)-3(h) plot PDOS decomposed onto 2p orbitals of each element type for a C/BN nanoribbon in Lattice #2 under various tensile strains corresponding to Fig. 2(d). In undeformed state [Fig. 3(e)], there are sizeable gaps in both the spin-up and spin-down band structures, with the same value about 0.35 eV. For both spin channels, the bottom of the conduction bands mainly originates from the 2p orbitals of carbon atoms, and so is for the top of the valence bands. As the applied tensile strain increases to 10% [Fig. 3(f)], for both the spin-up and spin-down states, the bottom of the conduction bands originating from boron atoms shifts toward the Fermi level. The top of the valence bands of the spin-down states, originating from carbon atoms, evolves closer to the Fermi level than that of the spin-up states. As a result, there is a gap about 0.2 eV for the spin-up states and a gap about 0.1 eV for the spin-down states. As the applied tensile strain increases to 12% [Fig. 3(g)], for the spin-up states, the gap between the bottom of the conduction bands and the top of valence bands increases to about 0.35 eV. However, for the spin-down states, the valance and conduction bands touch each other and the Fermi level falls within a zero-width gap. As the applied tensile strain further increases to 15% [Fig. 3(h)], for the spin-up states, the gap between the bottom of the conduction bands and the top of valence bands further increases to about 0.6 eV. For the spin-down states, the valance and conduction bands intersect with each other slightly at the vicinity of the Fermi level. The DOS results in Fig. 3(g) and Fig. 3(h) exhibit the typical SGS characteristics. The discussion above also indicates that the contribution from the 2p orbitals of carbon atoms at the ribbon edge is pivotal for the semiconductor-to-SGS transition of hybrid C/BN nanoribbons under tensile strains.

Emerging from our extensive DFT parametric studies are phase diagrams that quantitatively delineate the dependence of the electronic states of a hybrid C/BN nanoribbon on its lattice geometry and applied tensile strain. Figure 4(a) shows such a phase diagram for a C/BN nanoribbon in Lattice #1. It is clearly shown that half-metallicity can be achieved in a C/BN nanoribbon in Lattice #1 with a width greater than one BNR if a sufficient tensile strain is applied. The phase boundary defines a threshold tensile strain, above which a C/BN nanoribbon changes from a semiconductor to a HM. Such a threshold tensile strain decreases as the width of the nanoribbon (i.e., the number of BNRs in width direction) increases. Figure 4(b) plots the phase diagram for a C/BN nanoribbon in Lattice #2, which clearly demonstrates that SGS can be realized in a C/BN nanoribbon in Lattice #2 with a width greater than one BNR if a sufficient tensile strain is applied. The threshold tensile strain for the semiconductor-to-SGS transition decreases as the width of the nanoribbon increases.

## IV. Conclusions

In summary, we have explored the electronic properties of h-BN nanoribbons with one zigzag edge doped with a row of hexagonal carbon rings. We reveal that, when such a hybrid C/BN nanoribbon is subject to tension in its length direction, there exists a threshold tensile strain, above which the intrinsic semiconductive nanoribbon suddenly changes to a HM or a SGS, depending on the h-BN lattice structure. Since such a threshold tensile strain is below the elastic limit of the C/BN nanoribbons, such sharp semiconductor-to-HM or semiconductor-to-SGS transitions are reversible and thus programmable. These findings not only offer new solutions to the search of HM and SGS in material structures (e.g., 2D crystals) that are fundamentally distinct from those of existing solutions, but also promise a highly desirable but hard-to-achieve feature of programmable transition between different electronic states via strain engineering. Such a feature can potentially enable new design concepts of nanoelectronics. For example, by tuning the strain profile in hybrid 2D crystals via extrinsic regulations [28-30], it is possible to realize on-demand programming of the electronic states in various locations/components of an atomically thin electronic device. We therefore call for further theoretical and experimental investigations of these fertile opportunities.

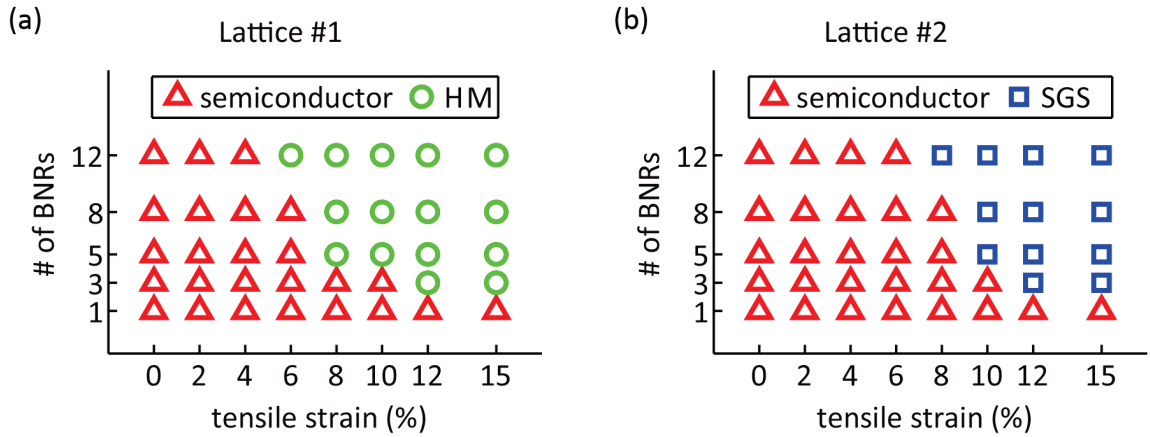

FIG. 4. Phase diagrams of the electronic states of a hybrid C/BN nanoribbon in the parametric space of ribbon width (# of BNRs) and applied tensile strain. The phase boundary delineates the threshold tensile strain for the onset of (a) semiconductor-to-HM transition in Lattice #1 and (b) semiconductor-to-SGS (spin-gapless semiconductor) transition in Lattice #2.

[1]   C. Felser, G. Fecher, and B. Balke, *Angew. Chem., Int. Ed.* **46**, 668 (2007).
[2]   X. Wang, *Phys. Rev. Lett.* **100**, 156404 (2008).
[3]   X. Wang, S. Dou, and C. Zhang, *NPG Asia Mater.* **2**, 31 (2010).
[4]   S. Ouardi, G. Fecher, C. Felser, and J. Kubler, *Phys. Rev. Lett.* **110**, 100401 (2013).
[5]   G. Xu, E. Liu, Y. Du, G. Li, G. Liu, W. Wang, and G. Wu, *Epl* **102**, 17007 (2013).
[6]   Y. Li, Z. Zhou, P. Shen, and Z. Chen, *ACS Nano* **3**, 1952 (2009).
[7]   Z. Wang, S. Jin, and F. Liu, *Phys. Rev. Lett.* **111**, 096803 (2013).
[8]   J. He, K. Chen, Z. Fan, L. Tang, and W. Hu, *Appl. Phys. Lett.* **97**, 193305 (2010).
[9]   Y. Fan, M. Zhao, X. Zhang, Z. Wang, T. He, H. Xia, and X. Liu, *J. Appl. Phys.* **110**, 034314 (2011).
[10]  J. Guan, W. Chen, Y. Li, G. Yu, Z. Shi, X. Huang, C. Sun, and Z. Chen, *Adv. Funct. Mater.* **23**, 1507 (2013).
[11]  J. Qi, X. Qian, L. Qi, J. Feng, D. Shi, and J. Li, *Nano Lett.* **12**, 1224 (2012).

[12]  L. Liu *et al.*, Science **343**, 163 (2014).
[13]  Y. Gao *et al.*, *Nano Lett.* **13**, 3439 (2013).
[14]  H. Xiao, C. He, C. Zhang, L. Sun, X. Peng, K. Zhang, and J. Zhong, *Phys. B* **407**, 4770 (2012).
[15]  J. Pruneda, *Phys. Rev. B* **81**, 161409 (2010).
[16]  W. Han, R. Kawakami, M. Gmitra, and J. Fabian, *Nat. Nanotechnol.* **9**, 794 (2014).
[17]  M. Guimaraes, J. van den Berg, I. Vera-Marun, P. Zomer, and B. van Wees, *Phys. Rev. B* **90**, 235428 (2014).
[18]  A. Geim and K. Novoselov, *Nat. Mater.* **6**, 183 (2007).
[19]  K. Novoselov, A. Geim, S. Morozov, D. Jiang, Y. Zhang, S. Dubonos, I. Grigorieva, and A. Firsov, Science **306**, 666 (2004).
[20]  K. Novoselov, A. Geim, S. Morozov, D. Jiang, M. Katsnelson, I. Grigorieva, S. Dubonos, and A. Firsov, Nature **438**, 197 (2005).
[21]  Y. Zhang, Y. Tan, H. Stormer, and P. Kim, Nature **438**, 201 (2005).
[22]  X. Li, X. Wang, L. Zhang, S. Lee, and H. Dai, Science **319**, 1229 (2008).
[23]  Y. Pan and Z. Yang, *Phys. Rev. B* **82**, 195308 (2010).
[24]  A. Castro Neto, F. Guinea, N. Peres, K. Novoselov, and A. Geim, *Rev. Mod. Phys.* **81**, 109 (2009).
[25]  F. Guinea, M. Katsnelson, and A. Geim, *Nat. Phys.* **6**, 30 (2010).
[26]  R. Martinez-Gordillo and M. Pruneda, *Phys. Rev. B* **91**, 045411 (2015).
[27]  J. Soler, E. Artacho, J. Gale, A. Garcia, J. Junquera, P. Ordejon, and D. Sanchez-Portal, *J. Phys.: Condens. Matter* **14**, 2745, (2002).
[28]  T. Li and Z. Zhang, *J. Phys. D: Appl. Phys.* **43**, 075303 (2010).
[29]  T. Li, *Simul. Mater. Sci. Eng.* **19**, 054005 (2011).
[30]  S. Kusminskiy, D. Campbell, A. Castro Neto, and F. Guinea, *Phys. Rev. B* **83**, 165405 (2011).